# On the Analysis and Generalization of Extended Visual Cryptography Schemes

DaoShun Wang[1,*], Feng Yi[1], Xiaobo Li[2], Ping Luo[1] and Yiqi Dai[1]

[1]Department of Computer Science and Technology, Tsinghua University, Beijing, 100084, China

[2]Department of Computing Science, University of Alberta, Edmonton, Alberta, Canada

**Abstract**

An Extended Visual Cryptography Scheme (EVCS) was proposed by Ateniese et al. [3] to protect a binary secret image with meaningful (innocent-looking) shares. This is implemented by concatenating an extended matrix to each basis matrix. The minimum size of the extended matrix was obtained from a hypergraph coloring model and the scheme was designed for binary images only [3]. In this paper, we give a more concise derivation for this matrix extension for color images. Furthermore, we present a ($k$, $n$) scheme to protect multiple color images with meaningful shares. This scheme is an extension of the ($n$, $n$) VCS for multiple binary images proposed in Droste scheme [2].

## 1. Introduction

Naor and Shamir [1] introduced the theory of Visual Cryptography Scheme (VCS) that conceals a secret image by copying the shares onto transparencies and giving to $n$ participants. In a ($k$, $n$) VCS, the secret can be recovered by viewing the overlapped transparencies associated to $k$ participants. No information can be obtained with $k$-1 or fewer shares. Each share can be either a random-looking image or a meaningful image (for example, a house or a dog, a tree, etc.). When we stack together the transparencies associated to a qualified group of participants, we get the desired secret image with no trace of the meaningful shares. With a forbidden group of shares, no information can be obtained. A (2, 2) scheme using meaningful shares was proposed in [1]. Droste[2] designed an algorithm to compute the bound of pixel expansion and relative difference

---

[*] Corresponding author.
E-mail address: daoshun@mail.tsinghua.edu.cn (DaoShun Wang), li@cs.ualberta.ca (Xiaobo Li)

for a (*k, n*)-VCS, and introduced a model to construct combinational (*n, n*)-VCS which shares more than one binary secret image among *n* transparencies with meaningful content. Ateniese et al. [3] proposed an extended VCS (EVCS) to protect a binary secret image with *n* meaningful shares. The secret image is encoded in the basis matrices, and the "meaningful" disguise images are encoded in the extended matrices that are concatenated to the basis matrices. The minimum size of the extended matrix is computed by using the hypergraph coloring method. For an extended VCS with a general access structure, they pointed out that it is an NP-hard problem to compute the chromatic number of a hypergraph. For a (*k, n*) access structure, the hypergraph is called the complete uniform hypergraph with the chromatic number (namely pixel expansion) of $\lceil n/(k-1) \rceil$.

While the above schemes were proposed to protect binary secret images, Verheul and Van Tilborg[4] introduced a theoretical construction for a (*k, n*)-VCS to protect a color secret image. Rijmen and Preneel [5] proposed a color (2, 2)-VCS with pixel expansion $m = 4$. Each share includes four colors, red, green, blue and white. Hou [6] stated that it is not appropriate to fill the shares with red, green, blue and white colors from both the additive model and the subtractive model of chromatology. A number of recent VCS's for gray-levels/color images using random-looking shares can be found in [7-9].

In summary, among the exist VCS's, most shares are random-looking image. A black and white (*k, n*) EVCS [3] with optimal matrix extension has been obtained, it is natural idea to construct a color VCS with innocent-looking shares, especially for multiple color images. And it is desirable to obtain a concise analysis of the matrix extension.

This paper will discuss the properties of the extended matrix collection, then give the minimum size of the extended matrix collection of a binary (*k, n*)-EVCS, propose a (*k, n*)-EVCS for single color images, and a (*k, n*) EVCS for multiple color images. The previous schemes (originally proposed for binary or grayscale or color images) using random-looking shares can be directly extended to generate innocent-looking shares using our "extended matrix collection" model.

This paper is organized as follows. Section 2 analyzes the matrix expansion in EVCS for binary images and presents a concise derivation for the minimum matrix size, and discusses the security issues of using innocent-looking shares. Based on Lemma 2.1 of Section 2, Section 3 constructs an EVCS for grayscale images/color images, and an EVCS for multiple color images is presented in Section 4. Section 5 concludes the paper.

## 2. Binary EVCS

Our proposed $(k, n)$-EVCS takes $n+1$ images $CI_1, \ldots, CI_n$, SI as input and generates $n$ shares $S_1, \ldots S_n$ that are modifications of the $n$ input pictures $CI_1, \ldots, CI_n$. The $(n+1)$-th secret image SI is reconstructed by overlapping any $k$ of the $n$ output shares $S_1, \ldots S_n$. We call the $n$ input images $CI_1, \ldots, CI_n$ cover images.

Here, we briefly review Naor and Shamir's visual cryptography scheme [1]. In this scheme, the secret image consists of a collection of black and white pixels and each pixel is subdivided into a collection of $m$ black and white sub-pixels in each of the $n$ shares. These subpixels are printed in close proximity to each other so that the human visual system averages their individual black and white contributions. The collection of sub-pixels can be represented by an $n \times m$ Boolean matrix $S = [s_{ij}]$, where the element $s_{ij}$ represents the $j$-th subpixel in the $i$-th share. A white subpixel is represented as 0, and a black subpixel is represented as 1. $s_{ij} = 1$ if and only if the $j$-th subpixel in the $i$-th share is black. The gray-level of the combined share by stacking shares $i_1, \ldots, i_r$ is proportional to the Hamming weight (the number of 1's in the vector $V$) $H(V)$ of the OR $m$-vector $V$ of the rows $i_1, \ldots, i_r$ in $S$. This gray-level is interpreted as black by the user's visual system if $H(V) \geq d$, and as white if $H(V) \leq d - \alpha \times m$ for some fixed threshold $1 \leq d \leq m$ and relative difference $\alpha > 0$.

The following definition is the formal definition for the binary VCS given in Ref.[1].

**Definition 2.1.**[1] A solution to the $k$ out of $n$ visual secret sharing scheme consists of two collections of $n \times m$ Boolean matrices $C_0$ and $C_1$. To share a white pixel, the dealer randomly chooses one of the matrices in $C_0$, and to share a black pixel, the dealer randomly chooses one of the matrices in $C_1$. The chosen matrix defines the color of the $m$ sub-pixels in each one of the $n$ transparencies. The solution is considered valid if the following three conditions are met.

1. For any $S$ in $C_0$, the OR $V$ of any $k$ of the $n$ rows satisfies $H(V) \leq d - \alpha \times m$.
2. For any $S$ in $C_1$, the OR $V$ of any $k$ of the $n$ rows satisfies $H(V) \geq d$.
3. For any subset $\{i_1, \ldots, i_q\}$ of $\{1, \ldots, n\}$ with $q < k$, the two collections of $q \times m$ matrices $D_t$ for $t \in \{0, 1\}$ obtained by restricting each $n \times m$ matrix in $C_t$ (where t = 0, 1) to rows $i_1, \ldots, i_q$ are indistinguishable in the sense that they contain the same matrices with the same frequencies.

In this definition parameter $m$ is called the pixel expansion, which refers to the number of subpixels representing a pixel in the secret image. The relative difference α refers to the difference in weight between combined shares that come from a white pixel and a black pixel in the secret image.

For a visual cryptography scheme to be valid, these three conditions must be met. The first two conditions of this definition are called contrast and the third condition is called security. Lemma 2.1 below discusses this condition in the situation of innocent-looking shares.

From the definition 2.1, a binary $(k, n)$-VCS can be realized by the two Boolean matrices $C_0$ and $C_1$. The collections $C_0$ ( resp. $C_1$) can be obtained by permuting the columns of the corresponding basis Boolean matrix $B_0$ (resp. $B_1$) in all possible ways. $B_0$ and $B_1$ are called basis matrices. In this paper, all of the constructions are realized by using basis matrices.

Let $A$ and $B$ be Boolean two matrices with the same number of rows. $A \circ B$ be the concatenation of the two matrices, and $H(V_B^t)$ be the Hamming weight (the number of 1's) of the OR $V$ of any $t$ rows in the matrix $B$.

**Lemma 2.1.** Let $B_0$ and $B_1$ be the basis matrices for a $(k, n)$-VCS with pixel expansion $m$ and relative difference $\alpha$. Let $A_0$ and $A_1$ be the distinct Boolean matrices with $n$ rows and $m_0$ columns. $B_0 \circ A_i$ and $B_1 \circ A_j$ are basis matrices for a $(k, n)$-VCS, $i, j = 0, 1$, then $H(V_{A_0}^k) = H(V_{A_1}^k)$ for any $k$ rows must be met. The pixel expansion is $m_E = m + m_0$ and relative difference is $\alpha_E = \alpha \times m / (m + m_0)$.

**Proof.** First, we verify the security of the new $(k, n)$-VCS for any Boolean matrices $A_0$ and $A_1$.

Because $B_0$ and $B_1$ are the basis matrices of a $(k, n)$-VCS, we get $H(V_{B_0}^q) = H(V_{B_1}^q)$ for $1 \leq q \leq k - 1$. For $B_0 \circ A_i$ and $B_1 \circ A_j$, $i, j = 0, 1$, if $i = j$ then the security of the scheme is satisfied.

In the following, we discuss the security of $B_0 \circ A_i$ and $B_1 \circ A_j$ when $i \neq j$.

Case 1. If $H(V_{A_0}^q) = H(V_{A_1}^q)$, we get $H(V_{B_0 \circ A_i}^q) = H(V_{B_1 \circ A_j}^q)$, then the security of the new scheme is guaranteed.

Case 2. If $H(V_{A_0}^q) \neq H(V_{A_1}^q)$, we get $H(V_{B_0 \circ A_i}^q) \neq H(V_{B_1 \circ A_j}^q)$ for $i \neq j$.

We notice that the matrices $A_0$ and $A_1$ are independent of the basis matrices $B_0$ and $B_1$. From $H(V_{B_0 \circ A_0}^q) \neq H(V_{B_1 \circ A_1}^q)$ and $H(V_{B_0 \circ A_1}^q) \neq H(V_{B_1 \circ A_0}^q)$, we can probably distinguish the matrices $A_0$ and $A_1$, but the two matrices have nothing to do with the basis matrices $B_0$ and $B_1$. Each $B_0 \circ A_0$, $B_1 \circ A_1$, $B_0 \circ A_1$, and $B_1 \circ A_0$ is determined either from the basis matrix $B_0$ with 1/2 probability or from the basis matrix $B_1$ with 1/2 probability. So the security of the $(k, n)$-VCS is ensured. Note that this is different from Condition 3 of Definition 2.1. Although the two Hamming weights mentioned above are not the same, $B_0 \circ A_i$ and $B_1 \circ A_j$ are still perfectly secure.

To sum up, in the new $(k, n)$-VCS, the collection of basis matrices $C_0$ and $C_1$ consists of all the matrices obtained by permuting the columns of $B_0 \circ A_i$ and $B_1 \circ A_j$, $i, j = 0, 1$. For any subset $\{i_1, \ldots, i_q\}$ of $\{1, \ldots, n\}$ with $q < k$, the two collections of $q \times m$ matrices obtained by restricting each $n \times m$ matrix in $C_0$ and $C_1$ to rows $i_1, \ldots, i_q$ are indistinguishable in the sense that they contain the same matrices with the same frequencies.

Next, we prove the contrast condition.

If $i = j$, it is clear that $B_0 \circ A_i$ and $B_1 \circ A_j$ satisfies the contrast condition. The proof for the case of $i \neq j$ is as follows.

Because $B_0$ and $B_1$ are basis matrices for a $(k, n)$-VCS, from the definition 2.1, the OR of any $k$ of $n$ rows in $B_0$ and $B_1$ satisfy $H(V_{B_0}^k) \leq d - \alpha \times m$. and $H(V_{B_1}^k) \geq d$. It is clear that $H(V_{B_1}^k) - H(V_{B_0}^k) \geq \alpha \times m$.

Assuming $H(V_{A_0}^k) = h_0$, $H(V_{A_1}^k) = h_1$, $T_0 = B_0 \circ A_0$ and $T_1 = B_1 \circ A_1$, the following formulas can be obtained:

$$H(V_{T_0}^k) = H(V_{B_0}^k) + H(V_{A_0}^k) \leq d - \alpha \times m + h_0$$

$$H(V_{T_1}^k) = H(V_{B_1}^k) + H(V_{A_1}^k) \geq d + h_1$$

Thus, $H(V_{T_1}^k) - H(V_{T_0}^k) \geq \alpha \times m + (h_1 - h_0)$.

If $h_1 - h_0 \geq 0$, the relative difference of the new $(k, n)$-VCS can be satisfied.

If $h_1 - h_0 < 0$, suppose $h_1 - h_0 = -1$, in the case $H(V_{T_1}^k) - H(V_{T_0}^k) = \alpha \times m - 1$.

Notice that the parameter $\alpha \times m \geq 1$. If the above formula satisfies the contrast condition of the VCS, namely $H(V_{T_1}^k) - H(V_{T_0}^k) \geq 1$; thus $\alpha \times m - 1 \geq 1$, we obtain $\alpha \times m \geq 2$. In this case, we would have to modify the original parameters $m$ and $\alpha$. We also know the parameters $m$ and $\alpha$ have been given in the scheme. It is impractical to modify the original parameters $m$ and $\alpha$ to construct a VCS directly, so $h_1 \geq h_0$ is obtained.

If $T_0 = B_0 \circ A_1$ and $T_1 = B_1 \circ A_0$, we get $H(V_{T_1}^k) - H(V_{T_0}^k) = \alpha \times m + (h_0 - h_1)$. By using a method

similar to the one above, we obtain $h_0 \geq h_1$.

From $h_1 \geq h_0$ and $h_0 \geq h_1$ we obtain $h_1 = h_0$, so the principle of contrast in the new scheme is satisfied.

Obviously, the pixel expansion is $m_E = m + m_0$.

The relative difference is

$$\alpha_E \times m_E = H(V_{T_1}^k) - H(V_{T_0}^k)$$

$$= H(V_{B_1}^k) + H(V_{A_i}^k) - (H(V_{B_0}^k) + H(V_{A_j}^k))$$

$$= H(V_{B_1}^k) - H(V_{B_0}^k) = \alpha \times m$$

Thus, we have $\alpha_E = \alpha \times m / m_E$.

Inspired by Lemma 3 in Ref.[2] and definition 3.1 in Ref.[3], we extended the lemma 3[2]. We call the Boolean matrices $A_0$ and $A_1$ extended matrices which satisfies $H(V_{A_0}^k) = H(V_{A_1}^k)$ for the $k$ of a ($k$, $n$)-VCS. The following is an example from example 3.2 of Ref. [3] to illustrate the lemma 2.1.

**Example 2.1.** In a (2, 2)-VCS with meaningful shares, the basis matrices are $T_c^{c_1,c_2}$, where $c, c_1, c_2 \in \{w, b\}$. Here $c$ represents the color of the pixel in the secret image. $c_1$ and $c_2$ represent the color of the pixels in the first and the second cover images. The $w$ pixel (resp. $b$ pixel) represents a white pixel (resp. black pixel).

$$T_w^{ww} = \begin{pmatrix} 1001 \\ 1010 \end{pmatrix} \text{ and } T_b^{ww} = \begin{pmatrix} 1001 \\ 0110 \end{pmatrix}$$

$$T_w^{wb} = \begin{pmatrix} 1001 \\ 1011 \end{pmatrix} \text{ and } T_b^{wb} = \begin{pmatrix} 1001 \\ 0111 \end{pmatrix}$$

$$T_w^{bw} = \begin{pmatrix} 1011 \\ 1010 \end{pmatrix} \text{ and } T_b^{bw} = \begin{pmatrix} 1011 \\ 0110 \end{pmatrix}$$

$$T_w^{bb} = \begin{pmatrix} 1011 \\ 1011 \end{pmatrix} \text{ and } T_b^{bb} = \begin{pmatrix} 1011 \\ 0111 \end{pmatrix}$$

Here, $T_c^{c_1,c_2} = B_c \circ A^{c_1 c_2}$, $c, c_1, c_2 = \{w, b\}$. $B_w = \begin{pmatrix} 10 \\ 10 \end{pmatrix}$ and $B_b = \begin{pmatrix} 10 \\ 01 \end{pmatrix}$ are basis matrices of a (2, 2)-VCS, $A^{ww} = \begin{pmatrix} 01 \\ 10 \end{pmatrix}$, $A^{wb} = \begin{pmatrix} 01 \\ 11 \end{pmatrix}$, $A^{bw} = \begin{pmatrix} 11 \\ 10 \end{pmatrix}$ and $A^{bb} = \begin{pmatrix} 11 \\ 11 \end{pmatrix}$.

$$T_w^{ww} = B_w \circ A^{ww}, \qquad T_b^{ww} = B_b \circ A^{ww}$$

$$T_w^{wb} = B_w \circ A^{wb}, \qquad T_b^{wb} = B_b \circ A^{wb}$$

$$T_w^{bw} = B_w \circ A^{bw}, \qquad T_b^{bw} = B_b \circ A^{bw}$$

$$T_w^{bb} = B_w \circ A^{bb}, \qquad T_b^{bb} = B_b \circ A^{bb}$$

The four different matrices $A^{ww}$, $A^{wb}$, $A^{bw}$ and $A^{bb}$ represent the four combinations of the two original pixels of the two cover images. Generally, it is denoted as $A = \begin{pmatrix} * & 1 \\ 1 & * \end{pmatrix}$, $* \in \{w, b\}$. $A^{ww}$, $A^{wb}$, $A^{bw}$ and $A^{bb}$ are called extended matrices, and $A = \begin{pmatrix} * & 1 \\ 1 & * \end{pmatrix}$ is called extended matrix collection. The $i$-th * represents the pixel in the $i$-th cover image. If it is black, * = $b$; If it is white, * = $w$. The * in the first and second row can take the same value or the different value.

As there are $2^n$ distinct combinations of black pixels and white pixels in the $n$ cover images, there are $2^n$ distinct extended matrices, all of which can be represented by an extended matrix collection $A$. By using the extended matrix collection we obtain a $(k, n)$-EVCS in which the $i$-th share shows the same content of the $i$-th cover image, $i = 1, \ldots, n$.

Some properties of the extended matrix collection $A$ are given as follows.

**Lemma 2.2.** In a $(k, n)$-EVCS, the extended matrix collection $A$ with minimum pixel expansion satisfies the following two basic conditions.

Condition 1  Each row has only one "*".

Condition 2  The number of *'s in each column is at most $k - 1$.

**Proof.** In a binary $(k, n)$-EVCS, the $i$-th row in $A$ represents the pixel in the $i$-th cover image. Thus one * is needed to represent the color of the pixel. If any row in $A$ has more than one *, the redundant *'s do not contribute to make the $i$-th share meaningful, so they can be deleted. Therefore each row in the extended matrix collection $A$ has only one *. The condition is obtained.

From the lemma 2.1, the OR of any $k$ rows in the extended matrix collection $A$ has the same Hamming weight. Therefore the number of *'s in each column is not more than $k - 1$.

For convenience, we assume that the relative difference of the reconstructed secret image is the same as that of the reconstructed cover images (or shares).

**Theorem 2.1.** Let $m$ (resp. $\alpha$) be the pixel expansion (reps. the relative difference) of a black and white $(k, n)$-VCS. Then we can construct a binary $(k, n)$-EVCS using the $(k, n)$-VCS and $n \times m_0$ extended matrix collection A. In this case

$$m_0 \geq \left\lceil \frac{n}{k-1} \right\rceil.$$

The pixel expansion $m_E = m + m_0$ and the relative difference $\alpha_E = \alpha \times m / m_E$ in the binary $(k, n)$-EVCS

**Proof.** From the condition 1 of the lemma 2.2, there exists only one * in each row, so the total number of *'s in $A$ is equal to $n$. From the condition 2 of the lemma 2.2, there exist at most $k - 1$ *'s in each column, so the total number of *'s in $A$ is not more than $m_0 \times (k - 1)$. The result is that $m_0 \times (k - 1) \geq n \times 1$; thus $m_0 \geq \left\lceil \frac{n}{k-1} \right\rceil$. From the lemma 2.1 we have $m_E = m + m_0$ and $\alpha_E = \alpha \times m / m_E$. ∎

Observe that Theorem 2.1 arrived at a conclusion that is the same as the one in [3] but involved a much simpler derivation. It is equivalent to Theorem 6.2 of [3] based on the hypergraph coloring model.

The following example illustrates the construction of the extended matrices collection.

**Example 2.2.** Construct the extended matrices collection $A$ for a binary $(3, 5)$-EVCS. From the theorem 2.1, the pixel expansion of $A$ is $m_0 \geq \left\lceil \frac{n}{k-1} \right\rceil$, namely $m_0 \geq 3$ and an example of the extended matrix collection is

$$A = \begin{pmatrix} * & 1 & 1 \\ * & 1 & 1 \\ 1 & * & 1 \\ 1 & * & 1 \\ 1 & 1 & * \end{pmatrix}$$

In the extended matrix collection $A$, each row has one *'s, thus it can show two different gray levels. The OR of any three rows has the same Hamming weight. Let $B_0$ and $B_1$ be the basis matrices of a binary $(3, 5)$-VCS. Thus $T_0 = B_0 \circ A$ and $T_1 = B_1 \circ A$ are the basis matrices of a binary $(3, 5)$-EVCS.

# 3. Color EVCS

## 3.1. EVCS for true color

The black and white VCS in [1] can be extended to design the color VCS (or grey-levels VCS). Verheul and Van Tilborg [4] first described this extended method. A color image is seen as an array of pixels, each of which is $k_0, k_1, \ldots, k_{c-1}$. Here $c$ is the number of color and $k_i$ is called the $i$-th color picture with color .Clearly, the gray-levels of a grayscale image can be viewed as different colors.Each pixel is divided into $m$ subpixels of color 0, …, $c$-1.These subpixels interrelate with each other in the following way.

When subpixels are put on top of each other and held to the light, one sees a " generalized "or, i.e. if all subpixels are of color $i$ then one sees light of color $i$, otherwise one sees no light at all (i.e. black). The "OR" operation of the two subpixels in Verheul and van Tilborg scheme is shown as following:

$$\text{color } i + \text{color } i = \text{color } i$$

$$\text{color } i + \text{color } j = \text{black, if } i \neq j$$

$$\text{color } i + \text{black} = \text{black}$$

$$\text{here } i = 0, \ldots, c\text{-1}$$

Now, we propose a color (k, n)-EVCS by using extended matrix collection. From the condition 2 of lemma 2.2 we have obtained that the number of *'s in each column is at most k − 1. This condition ensures the contrast of the reconstructed secret image. In fact, whatever the color of the * in an extended matrix collection, the condition is kept. Therefore, the following theorem is a direct result of the theorem 2.1 above.

**Theorem 3.1.** Suppose a color ($k$, $n$)-VCS with pixel expansion $m$, thus the corresponding color ($k$, $n$)-EVCS has the minimum pixel expansion $m+\left\lceil \dfrac{n}{k-1} \right\rceil$.

The following example illustrates the theorem 3.1, we use scheme in [8] to construct basis matrices $B_1, B_2$ and $B_3$ for a colored (2, 3)-VCS with three colors 1,2,3.

**Example 3.1.** A (2, 3)-EVCS with true color. Each original image has three colors, namely $c$ = 1, 2 and 3.

The basis matrices $S_0$ and $S_1$ are given for the black and white (2, 3)-VCS in [10]

$$S_0 = \begin{pmatrix} 0 & 1 & 1 \\ 0 & 1 & 1 \\ 0 & 1 & 1 \end{pmatrix}, S_1 = \begin{pmatrix} 0 & 1 & 1 \\ 1 & 0 & 1 \\ 1 & 1 & 0 \end{pmatrix}$$

Following is the three basis matrices $B_1, B_2, B_3$ for a colored (2, 3)-VCS constructed in [8].

$$B_1 = S_0^{1 \to k \atop 0 \to 1} \circ S_1^{1 \to k \atop 0 \to 2} \circ S_1^{1 \to k \atop 0 \to 3} = \begin{pmatrix} 1 & k & k & 2 & k & k & 3 & k & k \\ 1 & k & k & k & 2 & k & k & 3 & k \\ 1 & k & k & k & k & 2 & k & k & 3 \end{pmatrix}$$

While we delete all-$k$ column of matrix $B_1$, we obtain.

$$B_1 = \begin{pmatrix} 1 & 2 & 3 & k & k & k & k \\ 1 & k & k & 2 & 3 & k & k \\ 1 & k & k & k & k & 2 & 3 \end{pmatrix}$$

The other two basis matrices $B_2$ and $B_3$ are constructed as follows.

$$B_2 = \begin{pmatrix} 2 & 2 & 3 & k & k & k & k \\ 2 & k & k & 1 & 3 & k & k \\ 2 & k & k & k & k & 1 & 3 \end{pmatrix}$$

$$B_3 = \begin{pmatrix} 3 & 1 & 2 & k & k & k & k \\ 3 & k & k & 1 & 2 & k & k \\ 3 & k & k & k & k & 1 & 2 \end{pmatrix}$$

The extended matrix collection is:

$$A = \begin{pmatrix} * & k & k \\ k & * & k \\ k & k & * \end{pmatrix}$$

The * in the $i$-th row is equal to the color of the $i$-th cover pixel, $c = 1, 2, 3$. The basis matrices of the (2, 3)-EVCS are:

$$T_c = A \circ B_c$$

To encode an original pixel with color $c$, the basis matrix $T_c$ is chosen. The experimental results are shown in Appendix A.

### 3.2. Grayscale/color ($k, n$)-EVCS with mixture model

Our scheme supports the RGB model. We defined an RGB color palette as a set of RGB colors. The intensity of a primary color can be defined as the gray-level in the gray-level palette. A primary color will have an intensity range between 0 and 1, with 0 representing black and 1 representing the maximum possible intensity of that color. The RGB color palette is created from three gray-level palettes, which represent the intensity palettes for red, green and blue. Combining the members in the gray-level palettes in all possible ways creates the color palette. The VCS based on a color mixture model was first introduced by Rijmen and Preneel [5].

Assume the secret image has $g \geq 2$ distinct gray-levels from the 1-st (representing a white pixel) to the $g$-th (representing a black pixel). To express the scheme conveniently, we assume that the difference between two neighboring gray-levels is the same for each original image including the secret image and the cover images.

**Theorem 3.2.** Let $A_2$ be an extended matrix collection of a $(k, n)$-EVCS with binary cover images. Thus $A_g = \underbrace{A_2 \circ ... \circ A_2}_{g-1}$ is an extended matrix collection of a $(k, n)$-EVCS with at most $g$ gray-levels cover images. The number of columns in $A_g$ is $m_g = (g-1) \times \left\lceil \frac{n}{k-1} \right\rceil$.

**Proof.** In a gray-level VCS, the gray-level can be shown through the relative difference $\alpha$, which is related to the Hamming weight. Hence, different Hamming weights represent different gray-levels. Suppose $A_2$ has $m_2$ columns and the $i$-th row denoted by $r_2$. From the theorem 2.1, $m_2 \geq \left\lceil \frac{n}{k-1} \right\rceil$. From the condition 1 of the lemma 2.2, each row in $A_2$ has only one *; thus the Hamming weight of the row $r_2$ satisfies $H(r_2) \in \{m_2, m_2 - 1\}$.

Assume the $i$-th row in $A_g$ denoted by $r_g$.
$$H(r_g) = \underbrace{H(r_2) + ... + H(r_2)}_{g-1} \in \{(g-1) \times m_2, ..., (g-1) \times m_2 - (g-1)\}$$

Thus $r_g$ can show $g$ distinct gray-levels. The row $r_g$ is selected randomly from $A_g$, thus each row in $A_g$ can represent $g$ distinct gray-levels. Therefore $A_g$ is the extended matrix collection for a $(k, n)$-EVCS in which each cover image has at most $g$ gray-levels. The pixel expansion of $A_g$ is $m_g = (g-1) \times \left\lceil \frac{n}{k-1} \right\rceil$.

**Example 3.2.** This example illustrates the content given in the theorem 3.1. Based on the extended matrix collection $A_2$ of a $(3, 5)$-EVCS with black and white cover images, we can construct the extended matrix collection $A_3$ of a (3, 5)-EVCS with at most 3 gray-levels cover images.

$$A_2 = \begin{pmatrix} * & 1 & 1 \\ * & 1 & 1 \\ 1 & * & 1 \\ 1 & * & 1 \\ 1 & 1 & * \end{pmatrix}$$

$$A_3 = A_2 \circ A_2 = \begin{pmatrix} * & 1 & 1 \\ * & 1 & 1 \\ 1 & * & 1 \\ 1 & * & 1 \\ 1 & 1 & * \end{pmatrix} \circ \begin{pmatrix} * & 1 & 1 \\ * & 1 & 1 \\ 1 & * & 1 \\ 1 & * & 1 \\ 1 & 1 & * \end{pmatrix} = \begin{pmatrix} * & * & 1 & 1 & 1 & 1 \\ * & * & 1 & 1 & 1 & 1 \\ 1 & 1 & * & * & 1 & 1 \\ 1 & 1 & * & * & 1 & 1 \\ 1 & 1 & 1 & 1 & * & * \end{pmatrix}$$

In the extended matrix collection $A_3$, each row has two *'s, thus it can show three different gray levels. The OR of any three rows has the same Hamming weight.

The theorem 3.1 provides a construction method which can solve gray-level $(k, n)$-EVCS. In fact, the pixel expansion in the theorem 3.1 can be reduced furthermore. In the following theorem the optimal pixel expansion is proven.

**Theorem 3.3.** In a $g$ gray-levels $(k, n)$-VCS, let the pixel expansion be $m$ and the relative difference be $\alpha$. In the corresponding $(k, n)$-EVCS, suppose the $n$ distinct cover images have distinct gray-levels $g_1, \ldots, g_n$. Assume $A$ is the extended matrix collection with $m_0$ columns. Thus, we get $m_0 \geq \left\lceil \sum_{i=1}^{n} \frac{g_i - 1}{k - 1} \right\rceil$. Therefore, The pixel expansion is $m_E = m + m_0$ and the relative difference is $\alpha_E = \alpha \times m / m_E$.

**Proof.** In the extended matrix collection $A$, the $i$-th row needs $g_i - 1$ *'s to represent $g_i$ distinct gray-levels. Thus the number of *'s in $A$ is $\sum_{i=1}^{n}(g_i - 1)$. From the lemma 2.1, the extended matrix collection is a Boolean matrix collection, so the condition 2 of the lemma 2.2 must be satisfied. Namely, the number of *'s in each column is not more than $k - 1$. Thus, the number of *'s in $A$ is not more than $(k - 1) \times m_0$. It results that

$$\sum_{i=1}^{n}(g_i - 1) \leq (k - 1) \times m_0$$

$$m_0 \geq \left\lceil \sum_{i=1}^{n} \frac{g_i - 1}{k - 1} \right\rceil$$

From the lemma 2.1, we have $m_E = m + m_0$. Suppose the basis matrix of the $g$ gray-levels $(k, n)$-VCS is $B_j$ which represents the $j$-th gray-level, $j = 1, \ldots, g$. The relative difference $\alpha_E$ is:

$$\alpha_E \times m_E = H(V^k_{B_{j+1} \circ A}) - H(V^k_{B_j \circ A})$$

$$= H(V^k_{B_{j+1}}) + H(V^k_A) - (H(V^k_{B_j}) + H(V^k_A))$$

$$= H(V^k_{B_{j+1}}) - H(V^k_{B_j})$$

$$= \alpha \times m$$

Thus, we have $\alpha_E = \alpha \times m / m_E$.

In the theorem 3.2 above, the result of the theorem 2.1 can be obtained when $g_1 = \ldots = g_n = 2$.

**Example 3.3.** Suppose A is the extended matrix collection for a (3, 5)-EVCS, in which each cover image shows three distinct gray-levels.

$$A = \begin{pmatrix} * & * & 1 & 1 & 1 \\ 1 & * & * & 1 & 1 \\ 1 & 1 & * & * & 1 \\ 1 & 1 & 1 & * & * \\ * & 1 & 1 & 1 & * \end{pmatrix}$$

In the extended matrix collection $A$, each row has two *'s, thus it can show three different gray levels. Each column has two *'s, thus, the OR of any three rows is all 1's. Compared with the example 3.1, the pixel expansion is reduced.

Here we assume that the difference between two neighboring gray-level is the same for each original image including the secret images and the cover images. Based on the theorem 3.2 and the color mixture model, the following color VCS is obtained from a gray-level VCS.

**Theroem 3.4**. In a $c$ colors $(k, n)$-VCS, let the pixel expansion be $m^R$, $m^G$ and $m^B$, and the relative differences be $\alpha^R$, $\alpha^G$ and $\alpha^B$. Here, superscript R, G and B represent the red, green and blue components respectively. Suppose the corresponding $(k, n)$-EVCS has $n$ distinct cover images with distinct colors $c_1, \ldots, c_n$. Assume $A$ is the extended matrix collection with $m_0 = m_0^R + m_0^G + m_0^B$ columns. Thus, we get $m_0^R \geq \left\lceil \sum_{i=1}^{n} \frac{c_i^R - 1}{k-1} \right\rceil$, $m_0^G \geq \left\lceil \sum_{i=1}^{n} \frac{c_i^G - 1}{k-1} \right\rceil$ and $m_0^B \geq \left\lceil \sum_{i=1}^{n} \frac{c_i^B - 1}{k-1} \right\rceil$. Therefore, the pixel expansion is $m_E^R = m^R + m_0^R$, $m_E^G = m^G + m_0^G$ and $m_E^B = m^B + m_0^B$, the relative differences are $\alpha_E^R = \alpha^R \times m^R / m_E^R$, $\alpha_E^G = \alpha^G \times m^G / m_E^G$ and $\alpha_E^B = \alpha^B \times m^B / m_E^B$. The overall contrast difference is $\alpha^* = (\alpha_E^R \times m_E^R + \alpha_E^G \times m_E^G + \alpha_E^B \times m_E^B) / (m_E^R + m_E^G + m_E^B)$.

In general, the color model can be described by two equivalent models: the additive model and the subtractive model. The primary colors are red, green and blue in the additive model and cyan, magenta and yellow in the subtractive model.

The two color models satisfy the following relationships: $C = 255 - R$, $M = 255 - G$, $Y = 255 - B$. In the (R, G, B) representation, (0, 0, 0) represents full black and (255, 255, 255) represents full white. In the (C, M, Y) representation, (0, 0, 0) represents full white and (255, 255, 255) represents full black.

Here we will present an example to interpret the theorem 4.1.

**Example 3.5.** Construct the extended matrices collection $A_{RGB}$ for a (3, 5)-EVCS with eight colors cover images. From Example 3.1, the extended matrix collection for a black and white (3, 5)-EVCS is

$$A = \begin{pmatrix} * & 1 & 1 \\ * & 1 & 1 \\ 1 & * & 1 \\ 1 & * & 1 \\ 1 & 1 & * \end{pmatrix}$$

The above matrix $A$ is the basis matrix of each primary color components.

In the RGB model, red, green and blue represents the lightest color in the color mixing model for VCS; black represents the darkest color. Thus, the element 1 in the basis matrix is substituted with black; the element 0 is substituted with red, green and blue in the basis matrix of the red, green, blue components, respectively. The basis matrix of the color EVCS is $A_{RGB} = A_R \circ A_G \circ A_B$, in $A_R$, * ∈ {black, red}, in $A_G$, * ∈ {black, green}, in $A_B$, * ∈ {black, blue}.

In the CMY model, cyan, magenta and yellow represents the darkest color in the color mixing model for VCS; white represents the lightest color. Thus, the element 0 in the basis matrix is substituted with white; the element 1 is substituted with cyan, magenta and yellow in the basis matrix of the cyan, magenta, yellow components, respectively. The basis matrix of the color EVCS is $A_{CMY} = A_C \circ A_M \circ A_Y$, in $A_C$, $* \in \{\text{white, cyan}\}$, in $A_M$, $* \in \{\text{white, magenta}\}$, in $A_Y$, $* \in \{\text{white, yellow}\}$.

## 4. Multi Secret Images EVCS

*4.1. (k, n)-multi secret images EVCS with grey-levels or true colors*

Droste[2] has discussed the black and white multi secret images VCS which offers the possibility to give the stack of each combination of the $n$ shares different information without any hints of the resulting pictures when stacking further transparencies.

Let $P = \{1, \ldots, n\}$ be the participants set. A subset $S \subseteq 2^P \setminus \{\phi\}$, i.e. a set of non-empty subsets of $\{1, \ldots, n\}$, defines which combinations shall reveal a pictures, i.e. the stack of the transparencies $i_1, \ldots, i_q$ reveals one, if and only if $\{i_1, \ldots, i_q\} \in S$. Again the construction of the $n$ shares will be done pixel-wise, assuming that all the pictures have the same resolution in pixels. To construct the shares for each pixel, a matrix will be chosen of an appropriate multi-set, where this matrix determines the subpixels of the $n$ shares. As there are $|S|$ combinations of the shares to consider, there are $2^{|S|}$ different combinations of black and white pixels. To handle these $2^{|S|}$ combinations of pixels one has to deal with $2^{|S|}$ different multi-sets $C^T$ of matrices. They are indexed by subsets $T \subseteq S$, where the meaning of an index $T$ is the following: for every element $\{i_1, \ldots, i_q\} \in T$ the stack of the shares $i_1, \ldots, i_q$ shall appear black, while the other stacks shall appear white. So the Hamming weight of the OR of the rows of $i_1, \ldots, i_q$ of a matrix $B \in C^T$ has to be greater for $\{i_1, \ldots, i_q\} \in T$ than for $\{i_1, \ldots, i_q\} \notin T$.

**Definition 4.1.**[2] Let $S$ be a subset of $2^P \setminus \{\phi\}$. Multi-sets $C^T$ (for all $T \subseteq S$) of $n \times m$ Boolean matrices are called an S-extended $n$ out of $n$ secret sharing scheme, if the following three properties are met:

1. For all $\{i_1, \ldots, i_q\} \in S$, there is an $d(\{i_1, \ldots, i_q\}) \in N^+$ such that the Hamming weight of the OR of the rows $i_1, \ldots, i_q$ is at least $d(\{i_1, \ldots, i_q\})$ for all matrices of $C^T$ where $\{i_1, \ldots, i_q\} \in T$

2. For all $\{i_1, \ldots, i_q\} \in S$, there is an $\alpha(\{i_1, \ldots, i_q\}) \in R^+$ such that the Hamming weight of the

OR of the rows $i_1, \ldots, i_q$ is at most $d(\{i_1, \ldots, i_q\}) - \alpha(\{i_1, \ldots, i_q\}) \cdot m$ for all matrices of $C^T$ where $\{i_1, \ldots, i_q\} \notin T$.

3. For all $\{i_1, \ldots, i_q\} \subset \{1, \ldots, n\}$, the restrictions of the multi-sets $C^T$ to the rows $i_1, \ldots, i_q$ contain the same elements with the same frequencies for all $T$ which are the same when restricted to subsets of $\{i_1, \ldots, i_q\}$.

Following we construct meaningful multi secrets $(k, n)$-VCS (MEVCS) for color images by concatenating an extended matrix collection.

**Theorem 4.1.** Let the participants set be $P = \{1, \ldots, n\}$. In a c colors $(k, n)$-multi secret images VCS, $s = \sum_{l=k}^{n} \binom{n}{l}$ secret images are shared. Suppose the subset $P_i = \{r_1, \ldots, r_{p_i}\}$ will recover the $i$-th secret image with color $c_i$, $1 \leq i \leq s$, $k \leq p_i \leq n$. Suppose the pixel expansion is $m$ and the relative difference of the $i$-th recovered image is $\alpha_i$. In the corresponding $(k, n)$-MEVCS, suppose the $j$-th cover image has $d_j$ gray-levels, $1 \leq j \leq n$. Thus, the pixel expansion of the extended matrix collection satisfies $m_0 \geq \left\lceil \dfrac{n}{k-1} \right\rceil$. The pixel expansion of the $(k, n)$-MEVCS is $m_E = m + m_0$, and the relative difference of the $i$-th recovered image is $\alpha_{i,E} = \dfrac{\alpha_i \cdot m}{m_E}$, $1 \leq i \leq s$.

The following is an example of a (2, 3) –MEVCS with three colors.

**Example 4.1.** In a (2, 3) –MEVCS, $s = \sum_{l=2}^{3} \binom{3}{l} = 4$ secret images are shared. Assume each original image has three colors, $c = 1, 2, 3$. From the above theorem 4.1, we get the extended matrix collection:

$$A = \begin{pmatrix} * & k & k \\ k & * & k \\ k & k & * \end{pmatrix}$$

The * in the $i$-th row is equal to the color of the $i$-th cover pixel, $c = 1, 2, 3$.

The basis matrices of the (2, 3)–MEVCS is $T = A \circ B_1 \circ B_2 \circ B_3 \circ B_4$. Here, $B_1$, $B_2$, $B_3$ and $B_4$ are obtained from scheme of [8], $B_i \in \{B_{i,1}, B_{i,2}, B_{i,3}\}$.

$$B_{11} = \begin{pmatrix} 1 & 2 & k & 3 & k \\ 1 & k & 2 & k & 3 \\ k & k & k & k & k \end{pmatrix} \quad B_{12} = \begin{pmatrix} 2 & 1 & k & 3 & k \\ 2 & k & 1 & k & 3 \\ k & k & k & k & k \end{pmatrix} \quad B_{13} = \begin{pmatrix} 3 & 1 & k & 2 & k \\ 3 & k & 1 & k & 2 \\ k & k & k & k & k \end{pmatrix}$$

$$B_{21} = \begin{pmatrix} 1 & 2 & k & 3 & k \\ k & k & k & k & k \\ 1 & k & 2 & k & 3 \end{pmatrix} \quad B_{22} = \begin{pmatrix} 2 & 1 & k & 3 & k \\ k & k & k & k & k \\ 2 & k & 1 & k & 3 \end{pmatrix} \quad B_{23} = \begin{pmatrix} 3 & 1 & k & 2 & k \\ k & k & k & k & k \\ 3 & k & 1 & k & 2 \end{pmatrix}$$

$$B_{31} = \begin{pmatrix} k & k & k & k & k \\ 1 & 2 & k & 3 & k \\ 1 & k & 2 & k & 3 \end{pmatrix} \quad B_{32} = \begin{pmatrix} k & k & k & k & k \\ 2 & 1 & k & 3 & k \\ 2 & k & 1 & k & 3 \end{pmatrix} \quad B_{33} = \begin{pmatrix} k & k & k & k & k \\ 3 & 1 & k & 2 & k \\ 3 & k & 1 & k & 2 \end{pmatrix}$$

$$B_{41} = \begin{pmatrix} 1 & 2 & k & k & 3 & k & k \\ 1 & k & 2 & k & k & 3 & k \\ 1 & k & k & 2 & k & k & 3 \end{pmatrix} \quad B_{42} = \begin{pmatrix} 2 & 1 & k & k & 3 & k & k \\ 2 & k & 1 & k & k & 3 & k \\ 2 & k & k & 1 & k & k & 3 \end{pmatrix}$$

$$B_{43} = \begin{pmatrix} 3 & 1 & k & k & 2 & k & k \\ 3 & k & 1 & k & k & 2 & k \\ 3 & k & k & 1 & k & k & 2 \end{pmatrix}$$

Note that the pixel expansion is $m_E = 25$ and the relative difference in each recovered image are all 1/25, that is $\alpha_{1,E} = \alpha_{2,E} = \alpha_{3,E} = \alpha_{4,E} = 1/25$. The experimental results are given in appendix B.

*4.2. Grayscale (k, n)-MEVCS with mixture model*

In [7], Blundo et al. have obtained the pixel expansion of a grayscale $(k, k)$-VCS.

**Lemma 4.1.** [7] In a $g$ gray-levels secret images $(k, k)$-VCS, it holds that the pixel expansion is $m \geq (g-1) \times 2^{k-1}$.

The following results are promoted by Ref. [2].

A gray-level $(k, n)$-MEVCS is constructed by concatenating an extended matrix collection. Any $q \in \{k, k+1, \ldots, n\}$ shares will reconstruct a distinct secret image.

**Theorem 4.1.** Let the participants set be $P = \{1, \ldots, n\}$. In a gray-level $(k, n)$-multi secrets VCS,

$s = \sum_{l=k}^{n} \binom{n}{l}$ secret images are shared. Suppose the subset $P_i = \{r_1, ..., r_{p_i}\}$ will recover the $i$-th secret image with $g_i$ gray-levels, $1 \leq i \leq s$, $k \leq p_i \leq n$. The pixel expansion is $m = \sum_{i=1}^{s}[(g_i - 1) \times 2^{p_i - 1}]$ and the relative difference of the $i$-th recovered image is $\alpha_i$. In the corresponding $(k, n)$-MEVCS, suppose the $j$-th cover image has $h_j$ gray-levels, $1 \leq j \leq n$. Thus, the pixel expansion of the extended matrix collection satisfies $m_0 \geq \left\lceil \sum_{j=1}^{n} \frac{h_j - 1}{k - 1} \right\rceil$. The pixel expansion of the $(k, n)$-MEVCS is $m_E = m + m_0$, and the relative difference of the $i$-th recovered image is

$$\alpha_{i,E} = \frac{\alpha_i \cdot m}{m_E}, 1 \leq i \leq s.$$

**Proof.** For any $P_i = \{r_1, ..., r_{p_i}\}$, let $D_{it}$ be the basis matrices of a $(p_i, p_i)$-VCS with $g_i$ gray-levels, where $t = 1, ..., g$. From lemma 5.1, the number of columns in $D_{it}$ is $(g_i - 1) \cdot 2^{p_i - 1}$.

Next, we will prove that the $s = \sum_{l=k}^{n} \binom{n}{l}$ distinct secret images can be shared with pixel expansion $m \geq \sum_{i=1}^{s}[(g_i - 1) \times 2^{p_i - 1}]$.

Suppose $B_i[P_i]$ is a Boolean matrix with $(g_i - 1) \cdot 2^{p_i - 1}$ columns, and the 1-st row, ..., the $p_i$-th row in $B_i[P_i]$ are the $r_1$-th, ..., the $r_{p_i}$-th row in $B_i$ respectively. If the original pixel in the $i$-th secret image is the $g_{it}$-th gray-level, we have $B_i[P_i] = D_{it}$. All the rows in $B_i$ are all 1's except the $r_1$-th, ..., the $r_{p_i}$-th rows.

Now, we will prove the contrast condition is satisfied.

Let $B = B_1 \circ ... \circ B_s$ with $\sum_{i=1}^{s}(g_i - 1) \times 2^{p_i - 1}$ columns. For any $P_i = \{r_1, ..., r_{p_i}\}$, the OR of the $p_i$ rows $r_1, ..., r_{p_i}$ in $B$ has Hamming weight

$$H(V_B^{p_i}) = \sum_{j=1}^{s} H(V_{B_j}^{p_i}) = \sum_{\substack{j=1 \\ j \neq i}}^{s} H(V_{B_j}^{p_j}) + H(V_{B_i}^{p_i})$$

From the construction method and the security of $(p_j, p_j)$-VCS, $\sum_{\substack{j=1 \\ j \neq i}}^{s} H(V_{B_j}^{p_j})$ is a fixed number. Thus the principle of contrast is ensured.

Next, we will prove the security of the scheme.

The basis matrix of the $(k, n)$-MEVCS is $B = B_1 \circ \ldots \circ B_s$ in which the matrix $B_i$ are independent of the matrix $B_j$, $i \neq j$, $i, j \in \{1, \ldots, s\}$, and the information of the $i$-th secret image is only in $B_i$. For any matrix $B_l$, $1 \leq l \leq s$, selecting any $1 \leq q \leq p_l$ rows, no secret information of the $l$-th secret image can be obtained.

Select any $q$ rows in the basis matrix $B = B_1 \circ \ldots \circ B_s$, the security is discussed as fellows.

If $1 \leq q \leq k - 1$, the basis matrix $B$ provides no information of the $l$-th secret image.

If $k \leq q \leq p_l - 1$, the $i$-the secret image for $p_i < p_l$ can be recovered, but it provides no information of the $l$-th secret image.

Therefore, the security of the $l$-th secret image is ensured in the basis matrix $B = B_1 \circ \ldots \circ B_s$ for any $q$ rows, $k \leq q \leq p_l - 1$, $1 \leq l \leq s$. The $(k, n)$-MEVCS satisfies the security condition.

From theorem 3.2, the pixel expansion of the extended matrix collection satisfies $m_0 \geq \left\lceil \sum_{j=1}^{n} \frac{h_j - 1}{k - 1} \right\rceil$. From the lemma 2.1, the pixel expansion of the MEVCS is $m_E = m + m_0$ and the relative difference of the $i$-th recovered image is $\alpha_{i,E} = \frac{\alpha_i \cdot m}{m_E}$, $1 \leq i \leq s$.

**Corollary 4.1.** In a binary $(k, n)$-MEVCS, any $q \in \{k, k+1, \ldots, n\}$ shares will reconstruct a distinct secret image. The pixel expansion satisfies $m_E \geq \left\lceil \frac{n}{k-1} \right\rceil + \sum_{l=k}^{n} \binom{n}{l} \times 2^{l-1}$.

From the corollary 4.1, all the combination of shares can recover $2^n$-n-1 distinct secret images, the pixel expansion is $m_E \geq n + \sum_{l=2}^{n} \binom{n}{l} \times 2^{l-1}$.

Note that, when $k = 2$, the corollary 4.1 is the $(n, n)$ S-extended VCS in the Ref. [2].

*4.3. Color $(k, n)$-MEVCS with mixture model*

The following theorem proposes a color $(k, n)$-MEVCS. The pixel expansion and the contrast difference are also given.

**Theorem 4.2.** Let the participants set be $P = \{1, \ldots, n\}$. In a color $(k, n)$-multi secrets VCS, $s = \sum_{l=k}^{n} \binom{n}{l}$ secret images are shared. Suppose the subset $P_i = \{r_1, \ldots, r_{p_i}\}$ will recover the $i$-th secret image with $c_i$ colors, $1 \leq i \leq s$, $k \leq p_i \leq n$. The pixel expansions are $m^R$, $m^G$ and $m^B$. And the relative difference of the $i$-th recoverd image is $\alpha_i$. In the corresponding $(k, n)$-MEVCS, suppose the $j$-th cover image has $d_j$ colors, $1 \leq j \leq n$. Thus, the pixel expansion of the color $(k, n)$-MEVCS satisfies $m_E \geq \sum_{l=1}^{3} \left\{ \left\lceil \sum_{j=1}^{n} \frac{d_j^{(l)} - 1}{k - 1} \right\rceil + \sum_{i=1}^{s} \left[ (c_i^{(l)} - 1) \times 2^{p_i - 1} \right] \right\}$. The superscript $l = 1$, $l = 2$ and $l = 3$ represents the red, green and blue components respectively. Here, the relative difference in the $i$-th recovered image is $\alpha_{i,E}^R = \frac{\alpha_i^R \cdot m^R}{m_E^R}$, $\alpha_{i,E}^G = \frac{\alpha_i^G \cdot m^G}{m_E^G}$ and $\alpha_{i,E}^B = \frac{\alpha_i^B \cdot m^B}{m_E^B}$. The overall relative difference in the $i$-th recovered image is $\alpha_{i,E}^* = \frac{\alpha_{i,E}^R \cdot m_E^R + \alpha_{i,E}^G \cdot m_E^G + \alpha_{i,E}^B \cdot m_E^B}{m_E^R + m_E^G + m_E^B}$, $1 \leq i \leq s$.

## 5. Conclusions

We have analyzed the matrix concatenation in the Extended Visual Cryptography Scheme ([3]) and given a more concise derivation for this matrix extension for color images. Based on this more direct approach, we have extended the (n, n) VCS for multiple binary images ([2]) to a (k, n) scheme for multiple color images with meaningful shares. The feasibility of our scheme is also demonstrated by examples.

## Acknowledgements

This research was supported in part by the National Science Fund of the People's Republic of China under Grant No. 90304014 and was also supported in part by Canadian Natural Sciences and Engineering Research Council under Grant OGP9198.

# Appendix A. An example colored (2, 3)-EVCS

The example below demonstrates a (2, 3)-EVCS with three colors. The pixel expansion is m=10 and the subpixels are arranged in a 5x2 pattern. Parts (a)-(c) are cover images. Part (d) is the secret image. The shares are shown in Parts (e)-(g). And Parts (h)-(k) are the reconstructed images using two or three shares.

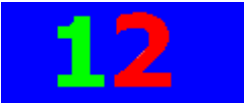

(a) Cover 1

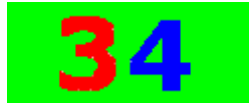

(b) Cover 2

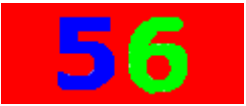

(c) Cover 3

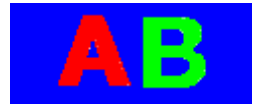

(d) Secret

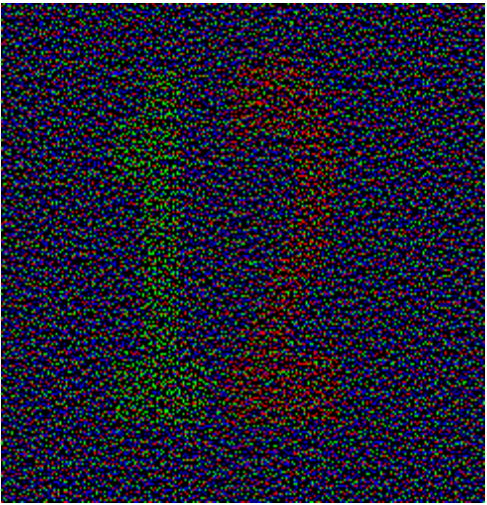

(e) Share 1

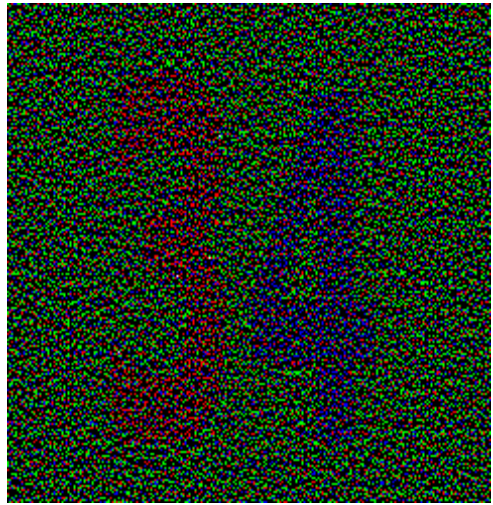

(f) Share 2

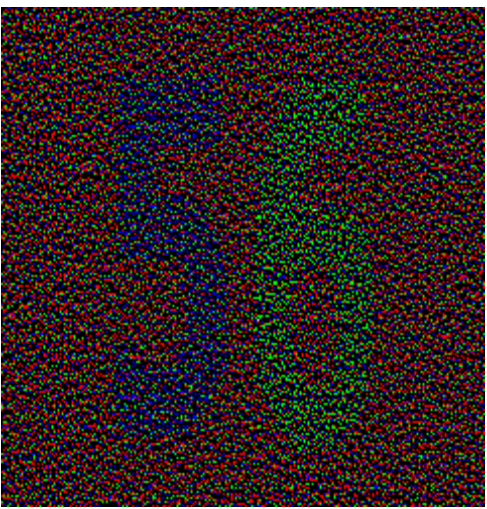

(g) Share 3

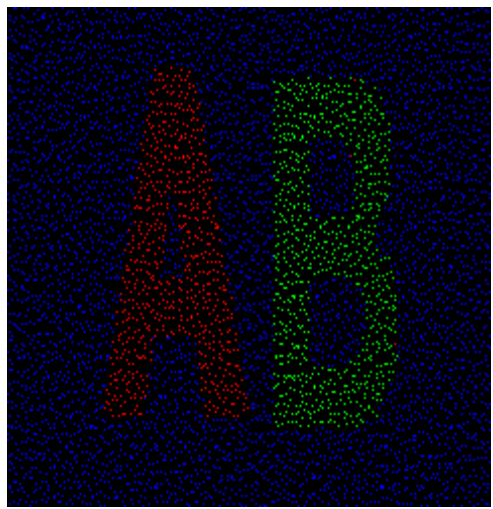

(h) Image of share 1 and 2

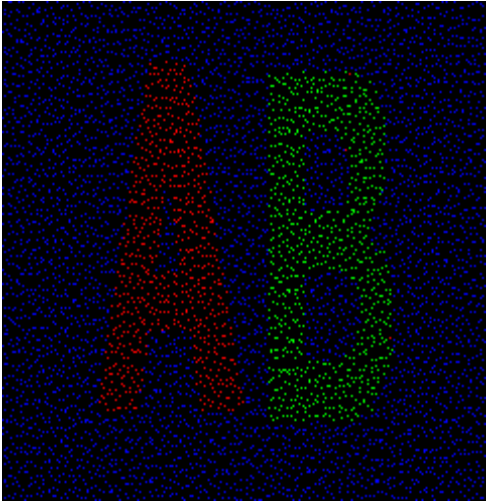

(i) Image of share 1 and 3

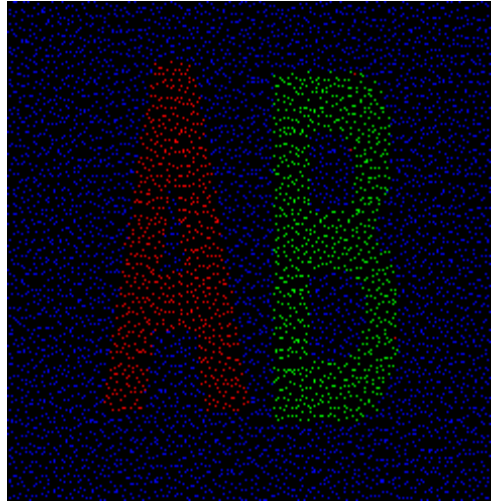

(j) Image of share 2 and 3

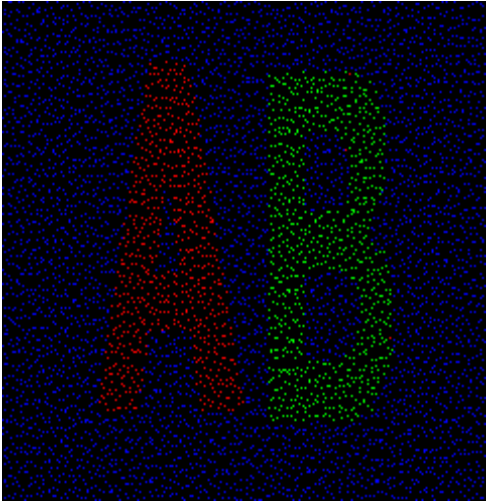

(k) Image of share1, 2 and 3

# Appendix B. An example colored (2, 3)-MEVCS

The example below demonstrates a (2, 3)-MEVCS with three colors. The pixel expansion is m=25 and the subpixels are arranged in a 5x5 pattern. Parts (a)-(c) are the cover images. Four difference secret images are show in Parts (d)-(g). The shares are shown in Parts (h)-(j). And Parts (k)-(n) are the reconstructed images.

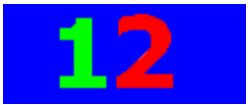 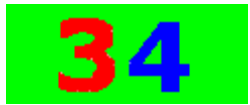 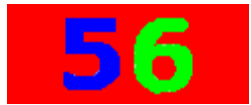

(a) Cove 1   (b) Cove 2   (c) Cove 3

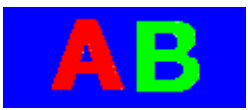 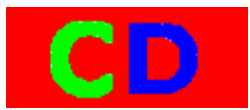 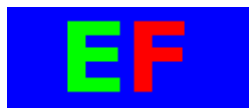 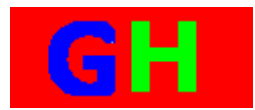

(d) Secret 1   (e) Secret 2   (f) Secret 3   (g) Secret 4

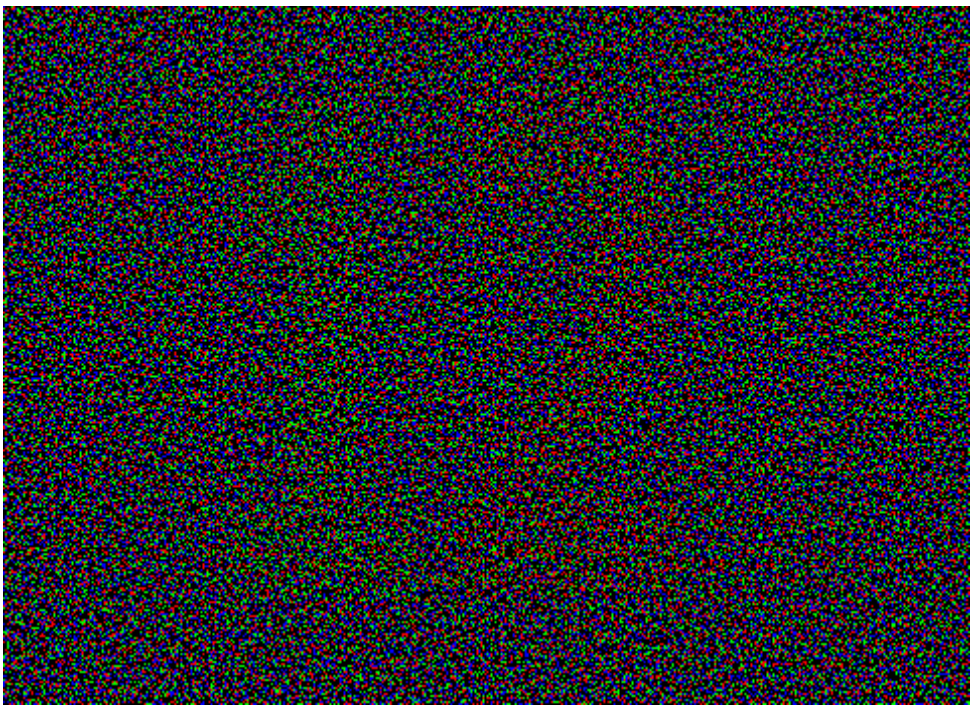

(h) Share 1

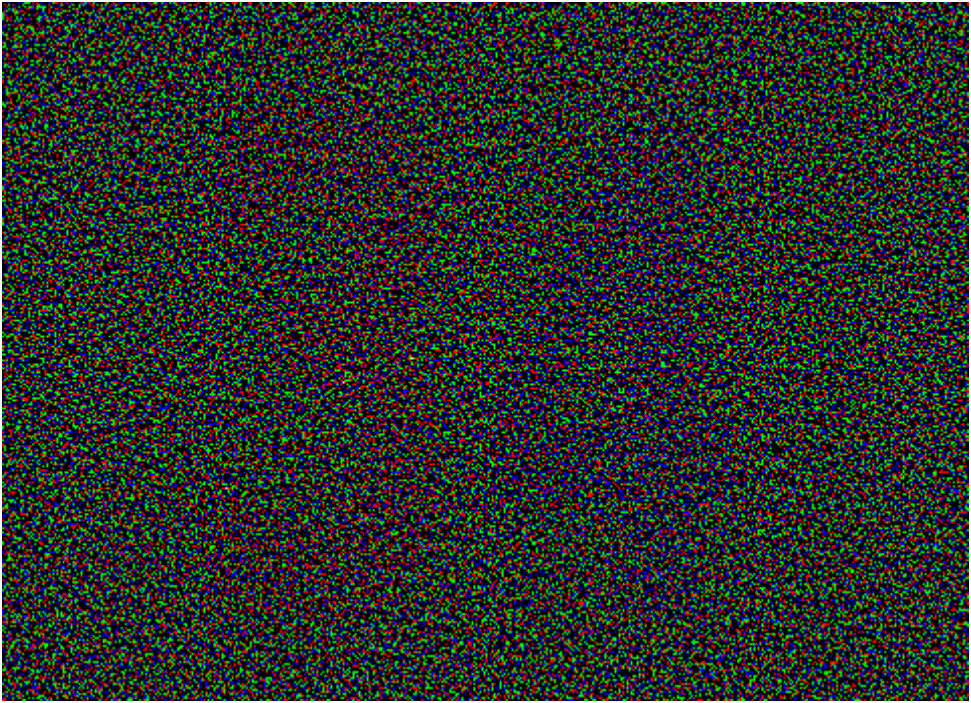

( i ) Share 2

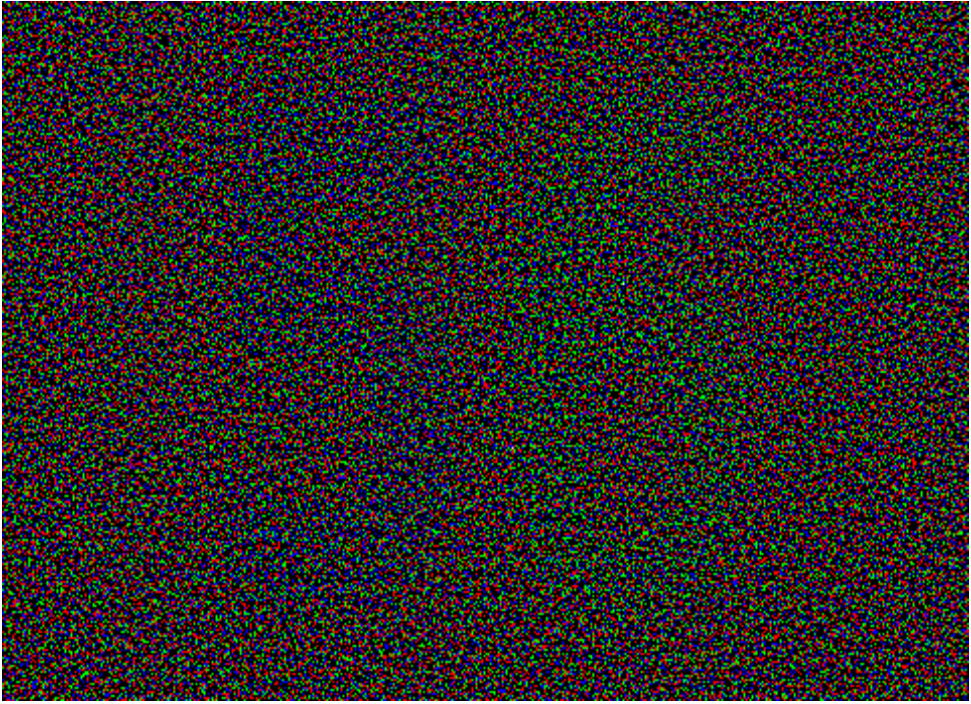

( j ) Share 3

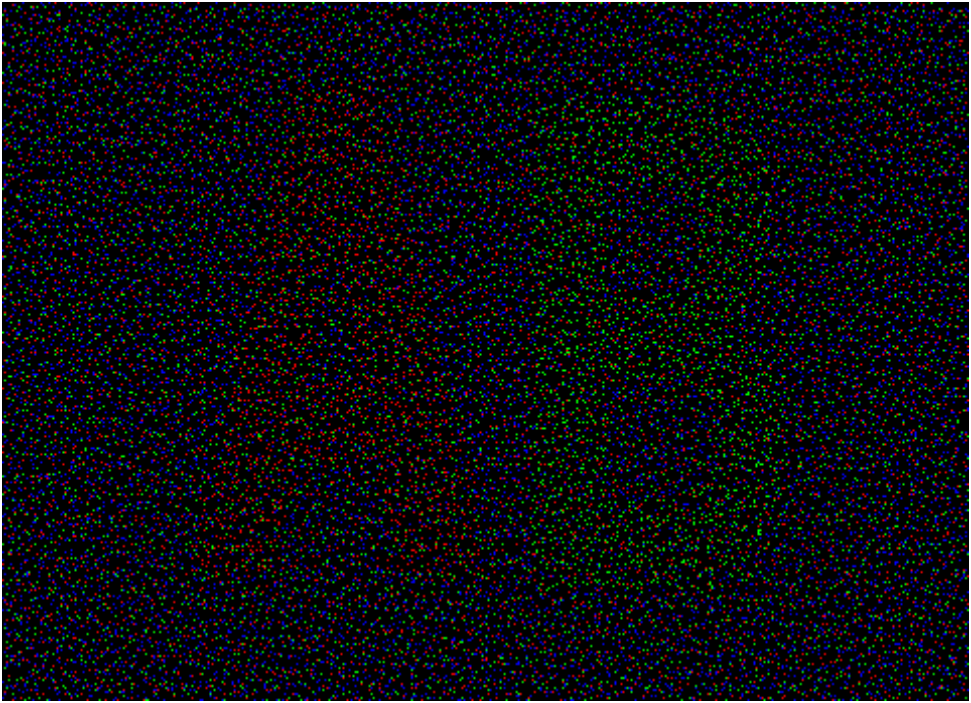

(k) Image of share 1 and 2

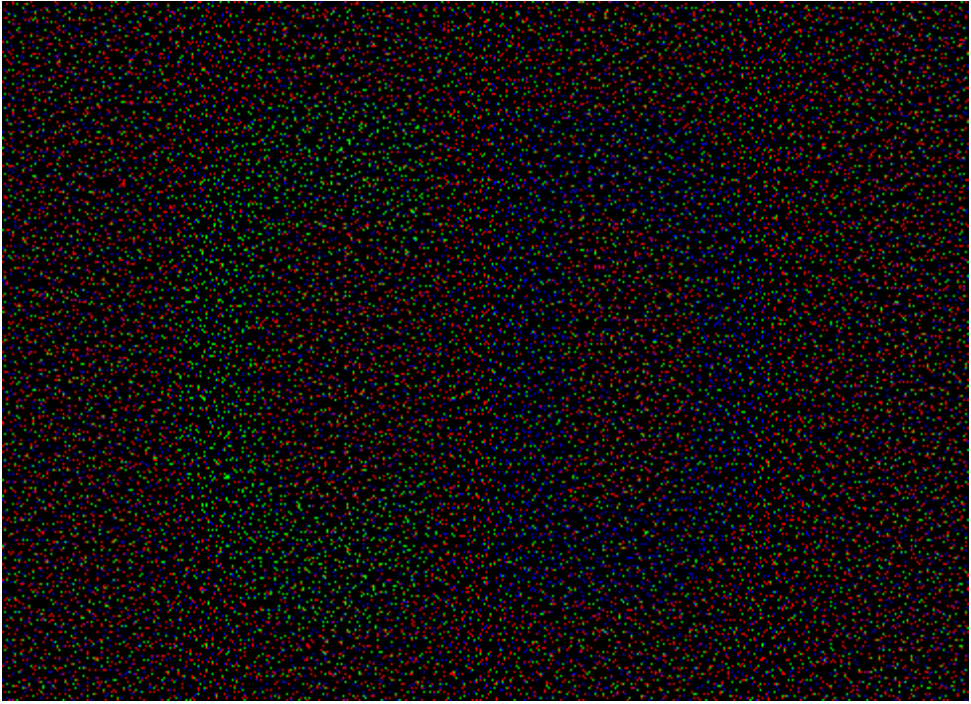

(l ) Image of share 1 and 3

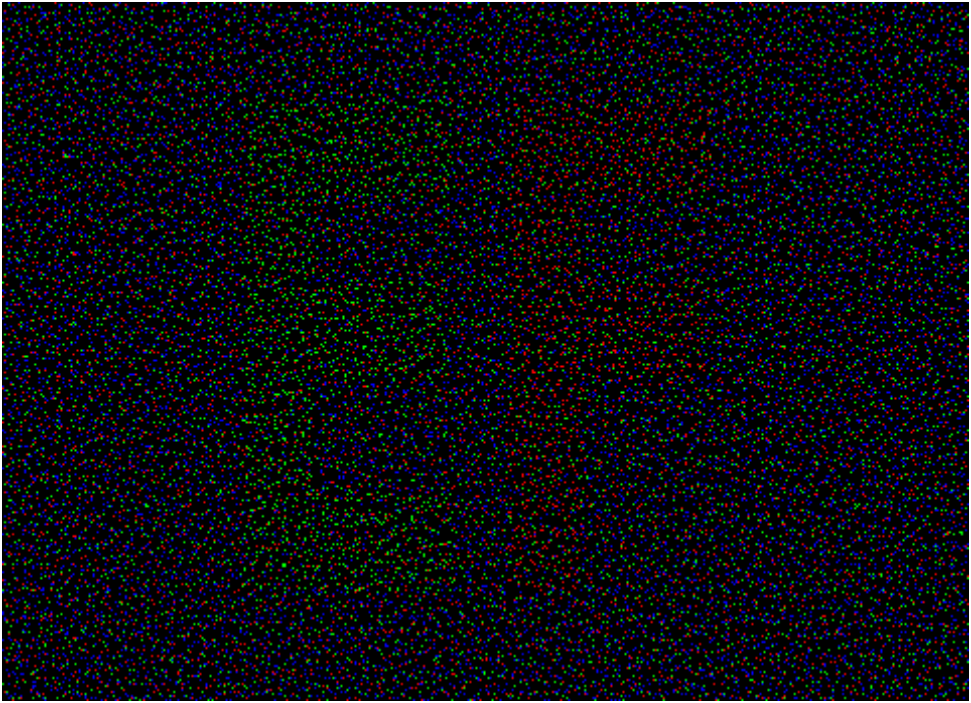

(m) Image of share 2 and 3

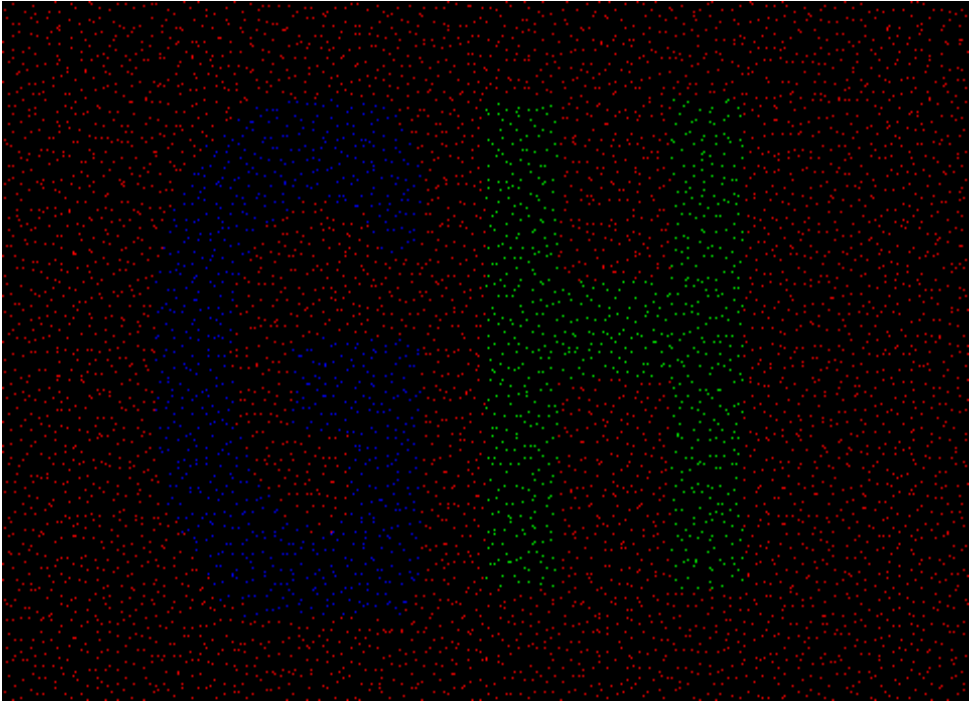

(n) Image of share 1, 2 and 3